\documentclass[sensors,communication,accept,moreauthors]{Definitions/mdpi} 
\usepackage{eurosym}

\firstpage{1} 
\pubvolume{1}
\issuenum{1}
\articlenumber{0}
\pubyear{2022}
\copyrightyear{2022}
\externaleditor{{Academic Editor: Wei Gao} 
 }
\datereceived{20 June 2022} 
\dateaccepted{3 August 2022} 
\datepublished{} 
\hreflink{https://doi.org/} 
\pdfoutput=1
\usepackage[final]{changes} 

\Title{Periodicity Intensity of the {24 h} 
 Circadian Rhythm in Newborn Calves Show Indicators of Herd Welfare}

\TitleCitation{Periodicity Intensity of the 24 h Circadian Rhythm in Newborn Calves Show Indicators of Herd Welfare}

\Author{{Victoria Rhodes}  $^{1}$\orcidC{}, Maureen Maguire $^{2}$, Meghana Shetty $^{2}$, Conor McAloon $^{1}$\orcidA{} and Alan F. Smeaton $^{2,3,}$*\orcidB{}}

\AuthorNames{Victoria Rhodes, Maureen Maguire, Meghana Shetty, Conor McAloon and Alan F. Smeaton}

\AuthorCitation{Rhodes, V.; Maguire, M.; Shetty, M.; McAloon, C.; Smeaton, A.F.}

\address{%
$^{1}$ \quad School of Veterinary Medicine, University College Dublin, Belfield, {D04 V1W8 Dublin, } 
 Ireland;   {vicki.rhodes@ucdconnect.ie (V.R.); conor.mcaloon@ucd.ie (C.M.)} 
\\ 
$^{2}$ \quad School of Computing, Dublin City University, Glasnevin, {Dublin 9,} 
 Ireland;  {\mbox{maureen.maguire47@mail.dcu.ie (M.M.);} meghana.shetty2@mail.dcu.ie (M.S.)} 
\\ 
$^{3}$ \quad Insight Centre for Data Analytics, Dublin City University, Glasnevin, {Dublin 9,} 
 Ireland
}

\corres{Correspondence: alan.smeaton@DCU.ie 
}

\abstract{
Circadian rhythms are a process of the sleep--wake
cycle that regulates the physical, mental and behavioural changes
in all living beings with a period of roughly 24 h. 
Wearable accelerometers are typically used in livestock applications to record animal movement from which we can estimate the activity type. Here, we use the overall movement recorded by accelerometers worn on the necks of newborn calves for a period of 8 weeks.  From the movement data, we calculate 24 h periodicity intensities corresponding to circadian rhythms, from a 7-day window that slides through up to 8-weeks of  data logging. The strength or intensity of the 24 h periodicity is computed at intervals as the calves become older, which is an indicator of individual calf welfare.
We observe that the intensities of these 24 h periodicities for individual calves, derived from movement data,  increase and decrease synchronously in a herd of 19 calves.
Our results show that external factors affecting the welfare of the herd    can be observed by processing and visualising movement data in this way and our method reveals insights that are not observable from movement data alone.
}

\keyword{animal welfare; activity monitoring; wearable sensors; data analytics; chronobiology; periodicity} 

\begin{document}

\section{Introduction}

Circadian rhythms play a crucial role in the lives of all {living things} and are vital for regulating sleep/wake cycles and feeding patterns. A~regular circadian rhythm can be linked to good health, wellbeing and a strong immune system.
Changes in circadian rhythm are known to occur under stressful conditions and thus an indicator of the intensity of the 24 h circadian rhythm may be a useful indicator of animal~welfare.

The regularity of the sleep-wake cycle can be measured using periodicity, the~phenomenon whereby an event or behaviour occurs at regular intervals over time. Periodicity identifies patterns in time series data that occur with regular periodic intervals.  The~detection of periodicity in time series {\deleted{is}can be} calculated {using a variety of techniques which were summarised in an extensive literature survey in~\cite{refinetti2007procedures}.  In~our work, we use}  power spectral density (PSD) \cite{nelson1981spurious} {\deleted{Using PSD, we can}as this allowed us to} determine the strength of repeating signals at {various} time frequencies, {though in the end, we focused on just the signal at the 24 h frequency, i.e., the~circadian rhythm}.

For our study, we used 24 h periodicity intensity, the~strength of the periodicity in sensor data at a frequency of 24 h, calculated within a sliding time window. A~strong periodicity indicates behaviour that is more regular and takes a value closer to 1. Weaker periodicity is closer to 0, {indicating} behaviour that is {more} random and~irregular. 

In a previous work, we have shown in human subjects how 24 h periodicity intensity correlates with the presence of blood markers which are known indicators of welfare~\cite{114b975b88414be6800d1ec6dc539ca6}. 
While {some} work has been conducted in humans, there has been a limited extension of the investigation of periodicity in commercial calves, though a recent systematic review of commercial sensors in precision livestock farming found a range of sensor technologies including  accelerometers, cameras, load cells, milk sensors, and~boluses~\cite{10.3389/fvets.2021.634338}. These are typically used to determine animal  activity~\cite{s21206816}, feeding and drinking behaviour~\cite{arcidiacono2017development}, and~overall welfare~\cite{jukan2017smart}.  There has been very little use of wearable sensors in calf welfare, especially in newborn calves, where the aim is to have high calf vitality, which can be defined as having the ``capacity to live and grow with physical and mental energy and strength'' \cite{murray2013}.
This is despite the knowledge that for dairy cows, the~consistency of behavioural patterns as measured by circadian rhythms is a characteristic of good health and welfare while changes to the regular patterns can be precursors of health issues
~\cite{10.3389/fanim.2022.839906,WAGNER202114}.

%
For the research reported herein, we  investigate the circadian periodicity of newborn calves and {how} the strength of that periodicity {changes throughout}  their first {6--}8 weeks.  Our method generates a historical overview of each calf and
{\deleted{while}} this analysis is useful to examine the development and welfare of each  individual calf. It is {also} possible that the regularity of the 24 h circadian rhythm {could be} a result of  a herd effect, {which is the focus herein}.
Many animals including cattle are behaviourally synchronised with herds, coordinating reactions to  external zeitgebers~\cite{conradt2003group}.
If we  determine a welfare indicator such as the periodicity intensity of the circadian rhythm, and~apply this to individual calves {\deleted{and to} in} a herd, then  the deviations of an individual calf from the herd {\deleted{welfare}} should stand out, {thus} allowing us to identify calves whose {individual} welfare is poorer than that of {the rest of} the~herd.

\section{Materials and~Methods}

\begin{figure}[H]
\includegraphics[width=10.5 cm]{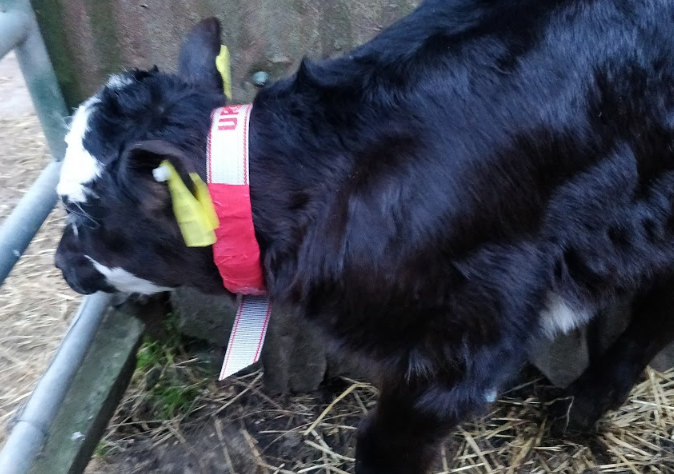}
\caption{Calf wearing a collar with Axivity AX3 sensor securely~affixed.\label{fig:cow}}
\end{figure}

Data were gathered from 24 dairy calves, each monitored for 60 days from birth, between~April and June 2019. 
{The calves were spring-born at a commercial dairy farm. 
They  were fed twice a day on bucket and teat, were bedded daily and had human contact during these activities.  Feeding times would have been at similar times to milking as they were fed whole milk after the automated milk feeders had  some technical issues.}  
The wearable sensors used to collect {\deleted{both}} data {\deleted{sets}} were Axivity AX3 sensors secured to a collar and placed around the necks of the newborn calves, as~shown in Figure~\ref{fig:cow}. The~collars were applied {\deleted{within 1 week} between 1 and 5 days} after the calves'~births.

The Axivity AX3 (Axivity, York, UK \url{https://axivity.com/}) is a popular {\deleted{device for wearable sensing research. It is a}} data logger with a built-in micro-electromechanical system {sensor in the form of} a three-axis accelerometer, which {enables it to monitor and} record movement levels against a real-time clock. The~expected battery life is 60 days between charges when {using our parameters which were} a sample rate of 12.5 Hz with 16-bit resolution and a range of + or $-$ 8 {G}. As~noted by Bucklet~et~al. \cite{buckley2020gait}, each sensor costs approximately EUR~120, measures 23.0 mm $\times$ 32.5 mm $\times$ \mbox{7.6 mm}, weighs 11 g and has {512~MB} of on-board memory for {data} storage{\deleted{of movement data}}. Data from the sensors can be accessed using the Open Movement (OmGUI) configuration and the analysis tool ({\url{https://github.com/digitalinteraction/openmovement/wiki/AX3-GUI/} (accessed on 15 June 2022)}), 
an open source application {\deleted{designed}} to set up, configure, download and visualise data from Axivity sensors. 
Although they {\deleted{sensors}}are waterproof to the IPx8 standard, they were wrapped in clingfilm, covered in masking tape and protective mesh and secured on the collar. {\deleted{Each was fully charged with the data sampling rate set to 12.5 Hz and recording turned on before being attached to collars.}} By the end of the collection period, each sensor{'s battery} had run out {\deleted{of battery}} but {\deleted{as the calves wore sensors for up to 8 weeks, this did not affect our data collection because}} our goal was to collect data for {at least} \mbox{6 weeks} {and that was achieved}.

Pre-processing data involved sampling x, y and~z axis values into signal vector magnitude (SVM), a~time-series independent of the sensor orientation and thus invariant to any movement of the collar {around the neck}, as we were interested in calves’ overall movement rather than movement in a particular direction. SVM shown below, also known as {{\it Amag}} 
\cite{RIABOFF2019104961}, is a useful metric for calculating movements which are not axis-specific and is used extensively in pre-processing raw accelerometer data~\cite{riaboff2022predicting}.  

\[ SVM = \sqrt{x^2 + y^2 + z^2}  \]

\noindent 
A Butterworth fourth-order band-pass filter~\cite{selesnick1998generalized} {with frequency in the range} 0.5 Hz and \mbox{20 Hz} was then applied to remove any white noise, and~negative values were {\deleted{handled by being}} converted into absolute values. The~final movement value used for analysing periodicity is the aggregated mean of SVM calculated over 60 epochs (seconds) which was chosen after we inspected different epochs between 1 min and up to 3600~s.

\begin{figure}[H]
\includegraphics[width=0.9\linewidth]{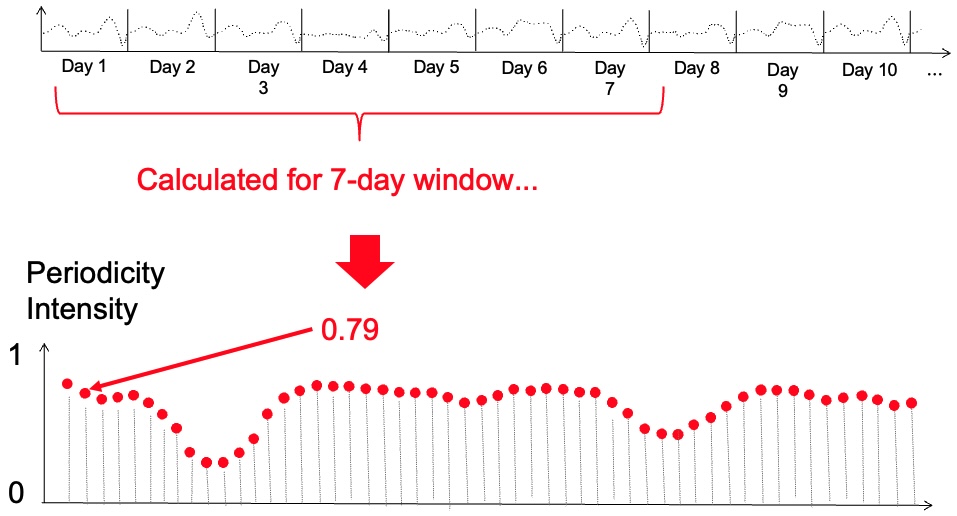}
\caption{Schematic for the calculation of periodicity intensity for a calf. Figure shows 10 days of SVM values (top part of the figure) with a 24 h periodicity intensity for a 7-day window calculated starting just after the start of recorded data and yielding a value of 0.79. This sliding window will shift forwards in time {and to the right on the schematic graph} by 15 min and re-calculate the 24 h periodicity intensity. This is repeated until the end of the data recording gives the time series plotted on the bottom half of the Figure, with~periodicity intensity values calculated at 15 min~intervals.\label{fig:window}}
\end{figure}

Determining the {intensity of} 24 h periodicity {\deleted{intensity}} over a {time series of up to 8 weeks}  {\deleted{entire 6-week time series}} would result in a single periodicity intensity for weeks of data, which would not provide any insights into the behaviour {during those weeks}. Therefore, we used a sliding window of \mbox{7 days} duration and determined the 24 h periodicity intensity for those 7 days, and~then shifted it by 15 min before repeating the process across the next 7 days of  data~\cite{hu2016periodicity}.  This time-lagged overlapping window across the entire data set for each calf provides intensity scores as schematically shown in Figure~\ref{fig:window}. The~{calculation} {\deleted{development}} of periodicity intensity over time was calculated for each~calf.

To look for synchronous impacts on periodicity {intensities} across the herd of 24 calves at {the same \deleted{given}} points in time, we aligned the time series of 24 h periodicity intensity values for each calf by  date and time and {we} present this as stacked line graphs{.} 
In a stacked line graph{,} the x axis represents  time from the beginning to the end of the  logging period, while the y axis is a stack of periodicity intensities for each of the calves, each {calf shown} in a different colour.
The top line of the overall graphs show the aggregated periodicities from {all} the calves and this will rise and fall during the logging period. 
Features to look for in the {stacked line} graph are where the overall graph rises or falls and where each or even many of the individual calf entries in the graph also rise or fall together; thus, the rises or falls in the overall graph can be attributed to many {rather than a small subset} of the calves' periodicity intensit{ies\deleted{y values}}.

\section{Results}

\begin{figure}[H]
\includegraphics[width=\linewidth]{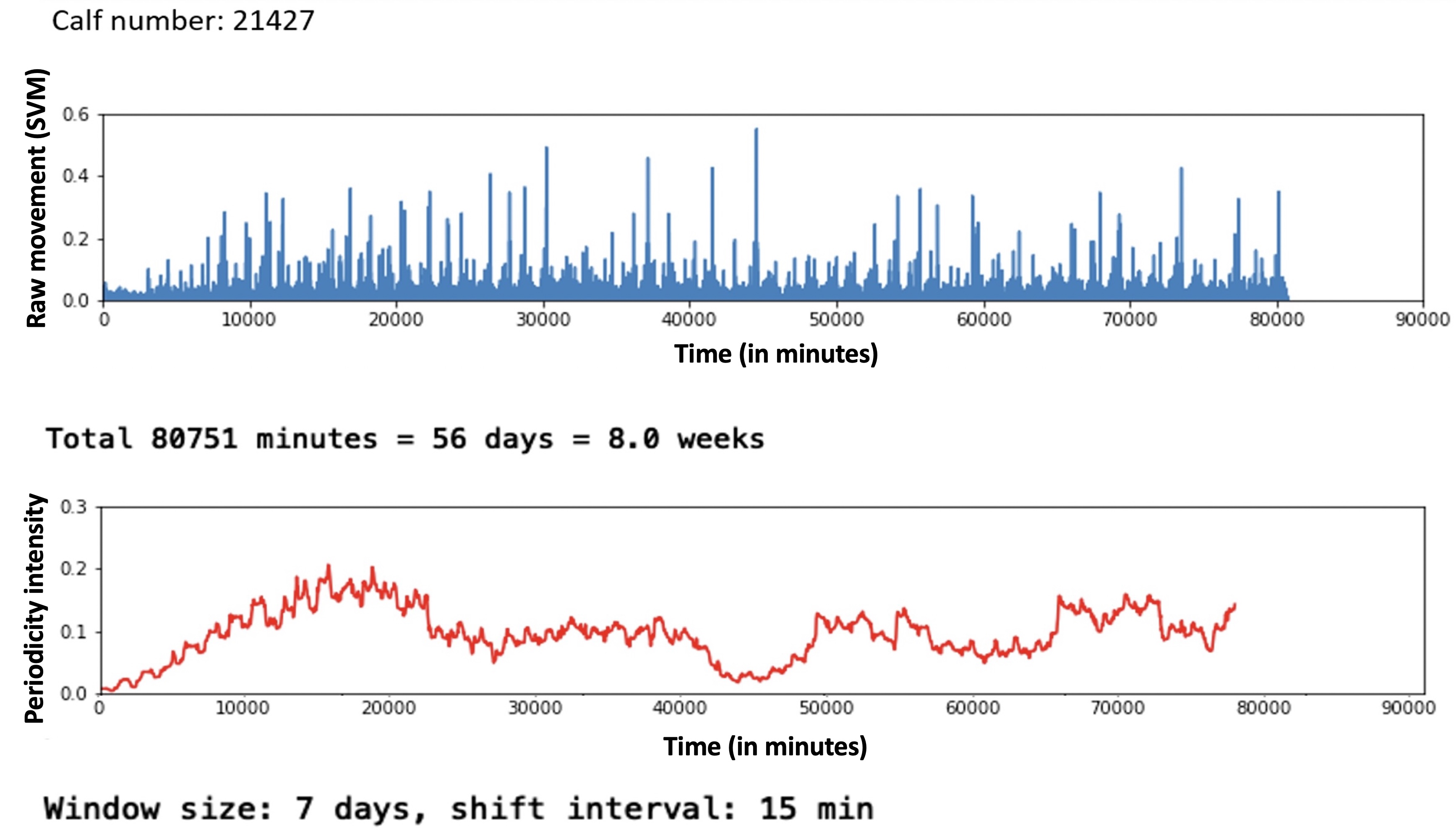}
\caption{{Raw data and periodicity intensity for} 
 calf 21427. The~top graph shows the SVM values computed from raw x--y--z accelerometer data as one value per minute. Peaks throughout the logging period correspond to calf activity during the day and examining raw values of activity levels suggests a reasonably regular circadian rhythm for this calf. The~bottom graph shows a time series of 24 h periodicity intensit{\deleted{y}ies}, each value calculated from a 7-day period and the {\deleted{calculated}calculation} is repeated every 15 min. This graph reveals a completely different picture for calf 21477 with a gradual acclimatisation to regular circadian rhythms for the first 14 days (c~19,000 min) followed by a drop-off reaching a low at approximately 31 days (c.~45,000 min) and then gradually recovering.
\label{fig:calf21427}}
\end{figure}

We analysed the 24 h periodicity intensities for the 24 calves to determine whether there were {simultaneous} changes observed across the herd{\deleted{ at the same time}}.  The~development of periodicity intensity for one of the calves {(number 21427)} is shown in Figure~\ref{fig:calf21427}. Periodicity intensity {is generally weak for this calf, averaging at approximately 0.1 throughout the logging period, whereas most of the other calves had higher values. The~explanation for this is that this calf is not as healthy as others in the herd. The~periodicity intensity} increased from {when the sensor was affixed shortly after} birth until 12--14 days {(c.~19,000 min)} of age after which there was a gradual drop-off with a low at approximately 31 days {(c.~45,000 min)} followed by a gradual recovery.  This pattern is not {\deleted{apparent}visible} in the raw SVM data {shown in the top graph in Figure~\ref{fig:calf21427}}, and while we do not know the {\deleted{case}cause} of this poor state of welfare at approximately day 31, it is useful to know whether this was systematic across the herd or {\deleted{individual}specific} to this particular~calf.

Although the 24 calves had different dates of birth, all calves in the herd were contributing accelerometer data by 5 April. Analysing data from this date onward{s} enables us to combine data from older and younger calves. By~this date, 5 calves were already \mbox{5 weeks} old (born on 2 March), 4 were already about 3 weeks old (those born 14 March), 4 more were 2 weeks old (born on 21 March) and 11 were newborns (born on 5 April). 

\begin{figure}[H]
\includegraphics[width=\linewidth]{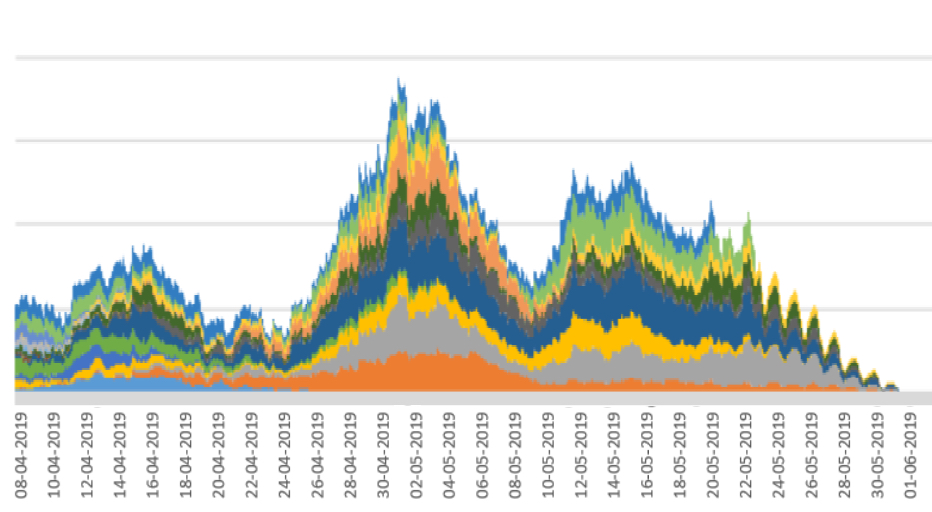}
\caption{Stacked line graph for aligned periodicity intensities for 19 contributing calves. Each colour represents a different~calf. \label{fig:stack}}
\end{figure}

Among the 24 calves, 5 were outliers and were eliminated from further  analysis as they had a large amount missing data due to their collars falling off and not being re-attached quickly enough. Such gaps in the recording of movement {\deleted{will}for an individual calf would} disrupt the calculation of the shifting 7-day windows of periodicity intensity.  The \mbox{19 remaining} calves were used to generate the stacked line graph shown in Figure~\ref{fig:stack}, where each colour represents a different calf and the analysis is applied to more than 1100 days of {\deleted{analysis}data gathering}. The~x axis in this figure represents dates. End dates are not consistent across individual calves due to the fact that {either a sensor's battery ran out 
or} movement sensors {with working batteries} were removed from calves at different~times following the data collection period. 

\vspace{-9pt}

In the analysis of the herd of calves in Figure~\ref{fig:stack}, we see that periodicity intensity {for the whole herd} increased rapidly  beginning on 24 April, rising up steadily  until about  3 May. Directly following this, there was a clear drop in 24 h periodicity intensity across the herd, indicating a synchronous dis-improvement in welfare, which reached its low point after about 1 week, around 9 May. This was followed by a return to better welfare, although~{\deleted{nit}not} as high as at 2 May, and~at around 24 May, individual calf data collection started to end as the accelerometers were removed from calves {or the battery ran out}. Since  periodicity intensity values are based on a 7-day sliding window, there will be a lag of some days between a stressful event  occurring at a single point in time  which could have caused this dis-improvement and evidence of it manifesting in the 7-day periodicity intensity window values.  Thus, {the} drop-off, {which reached its minimum point} around 9 May, {was most likely caused by a stressful or traumatic event at around 2 May.}

\begin{figure}[H]
\includegraphics[width=\linewidth]{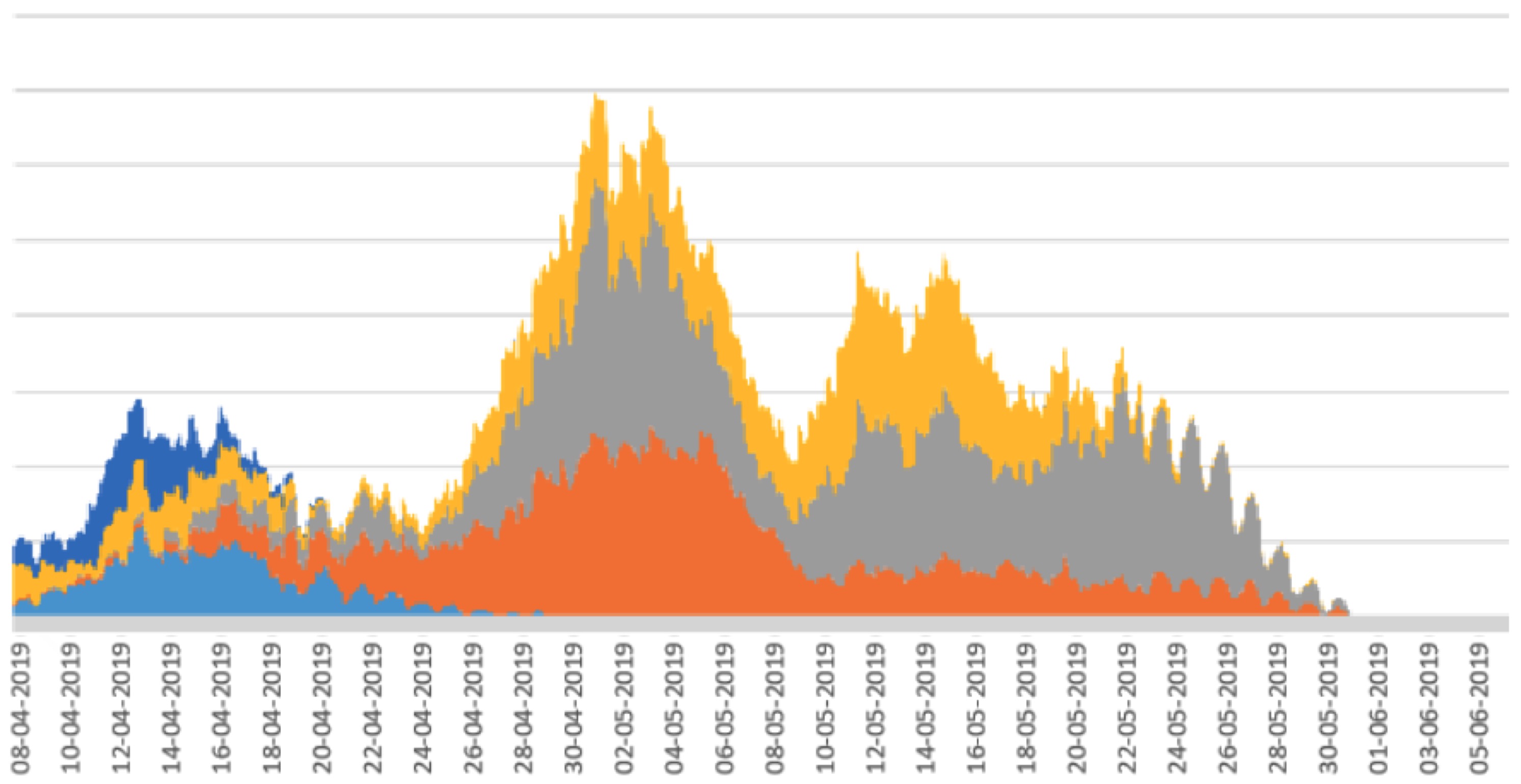}
\caption{Stacked graph for aligned periodicity intensities for calves 1--5. Each colour represents a different~calf. \label{fig:stack1}}
\end{figure}

\begin{figure}[H]
\includegraphics[width=\linewidth]{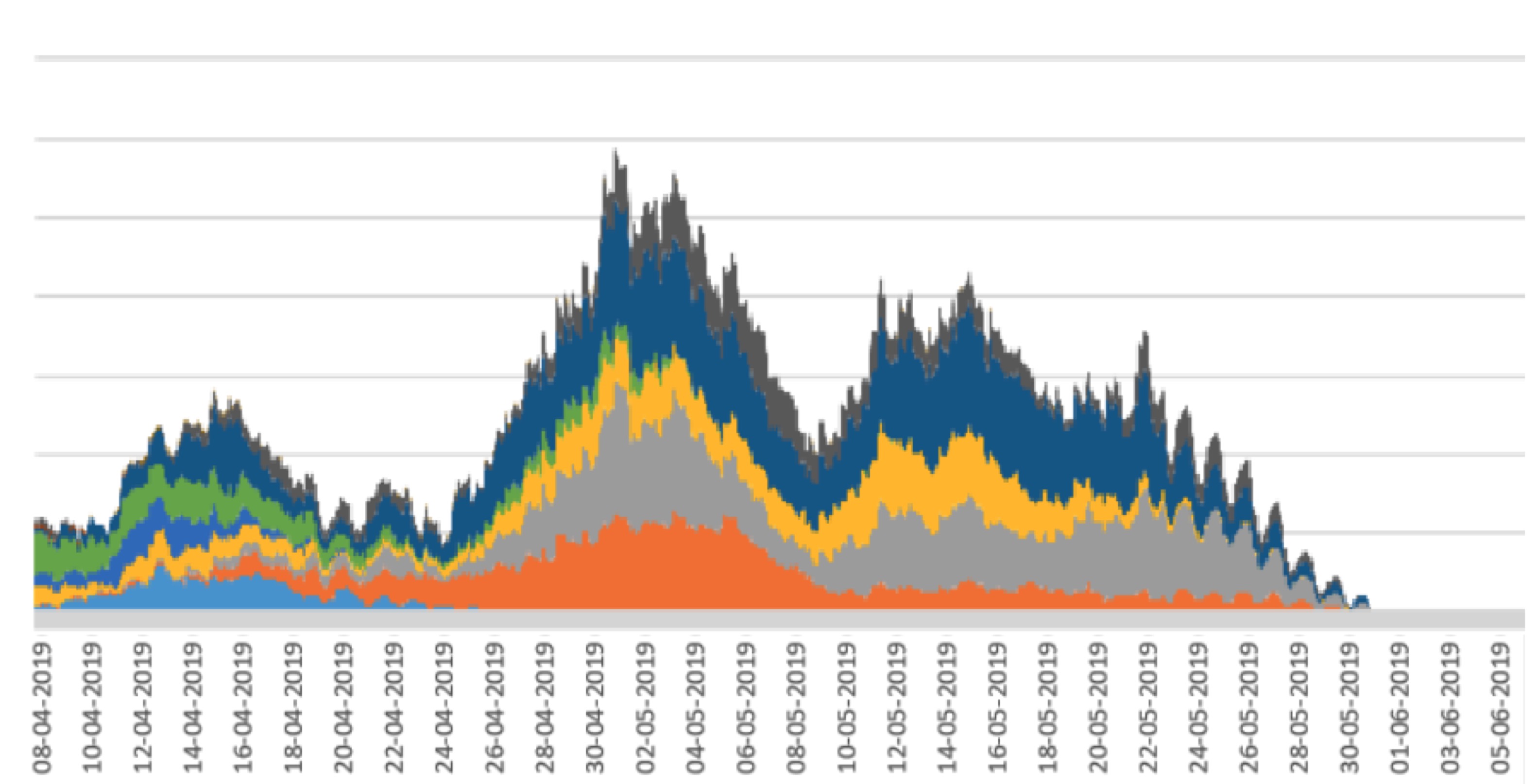}
\caption{Stacked line graph for aligned periodicity intensities for calves 1--10. Each colour represents a different~calf. \label{fig:stack2}}
\end{figure}

\begin{figure}[H]
\includegraphics[width=\linewidth]{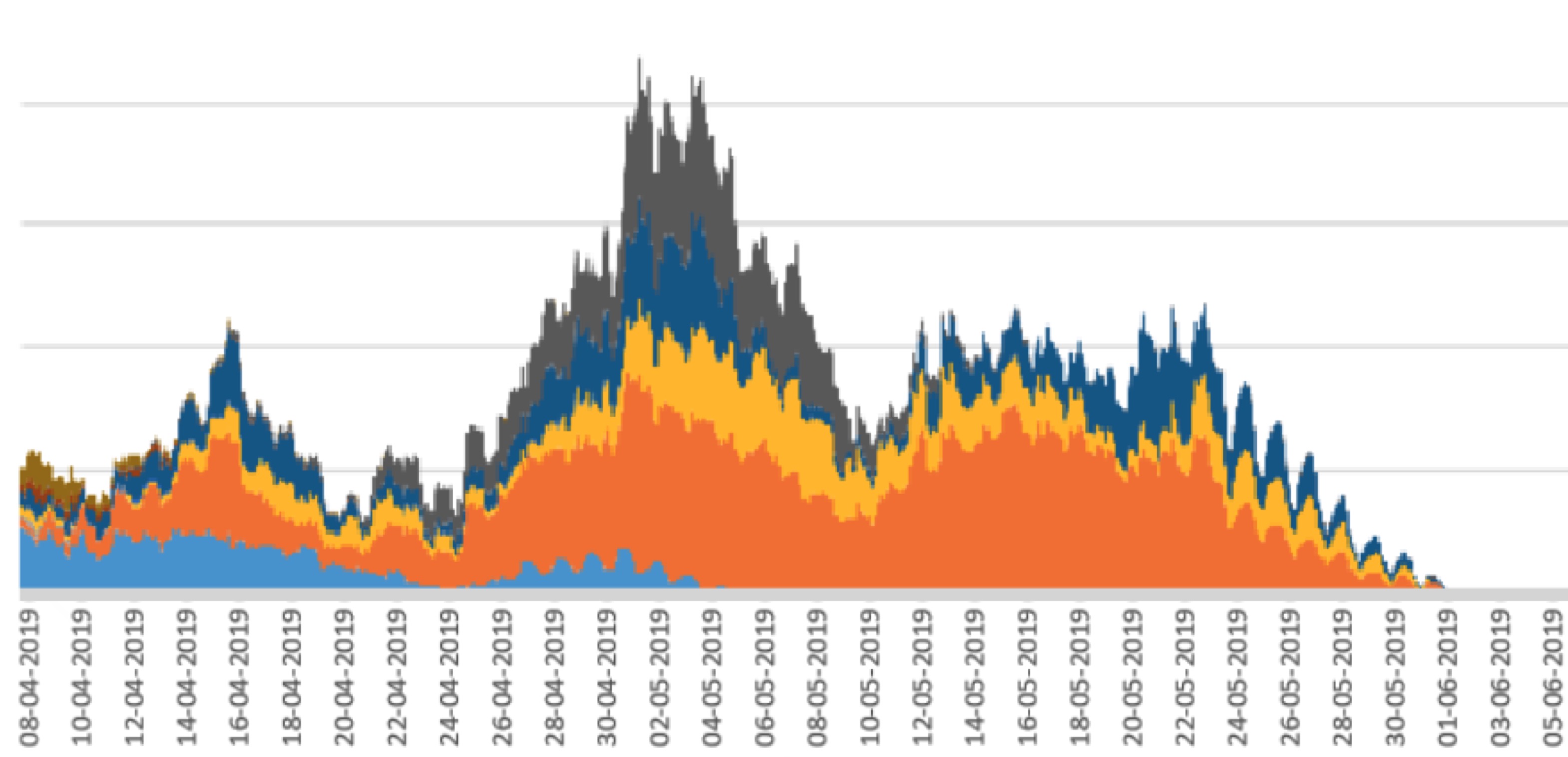}
\caption{Stacked line graph for aligned periodicity intensities for calves 6--15. Each colour represents a different~calf. \label{fig:stack3}}
\end{figure}

\begin{figure}[H]
\includegraphics[width=\linewidth]{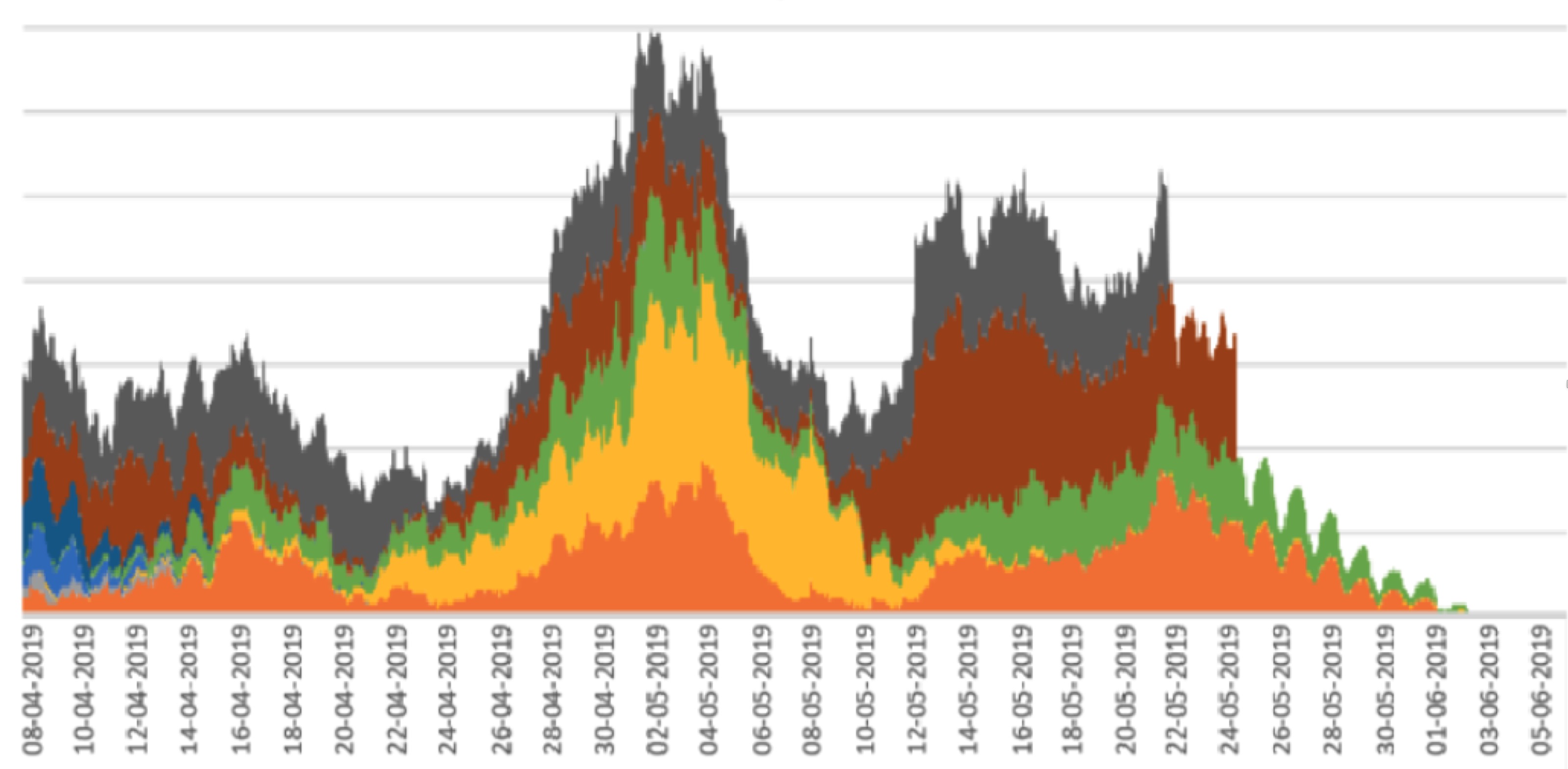}
\caption{Stacked line graph for aligned periodicity intensities for calves 11--19. Each colour represents a different~calf. \label{fig:stack4}}
\end{figure}

\begin{figure}[H]
\includegraphics[width=\linewidth]{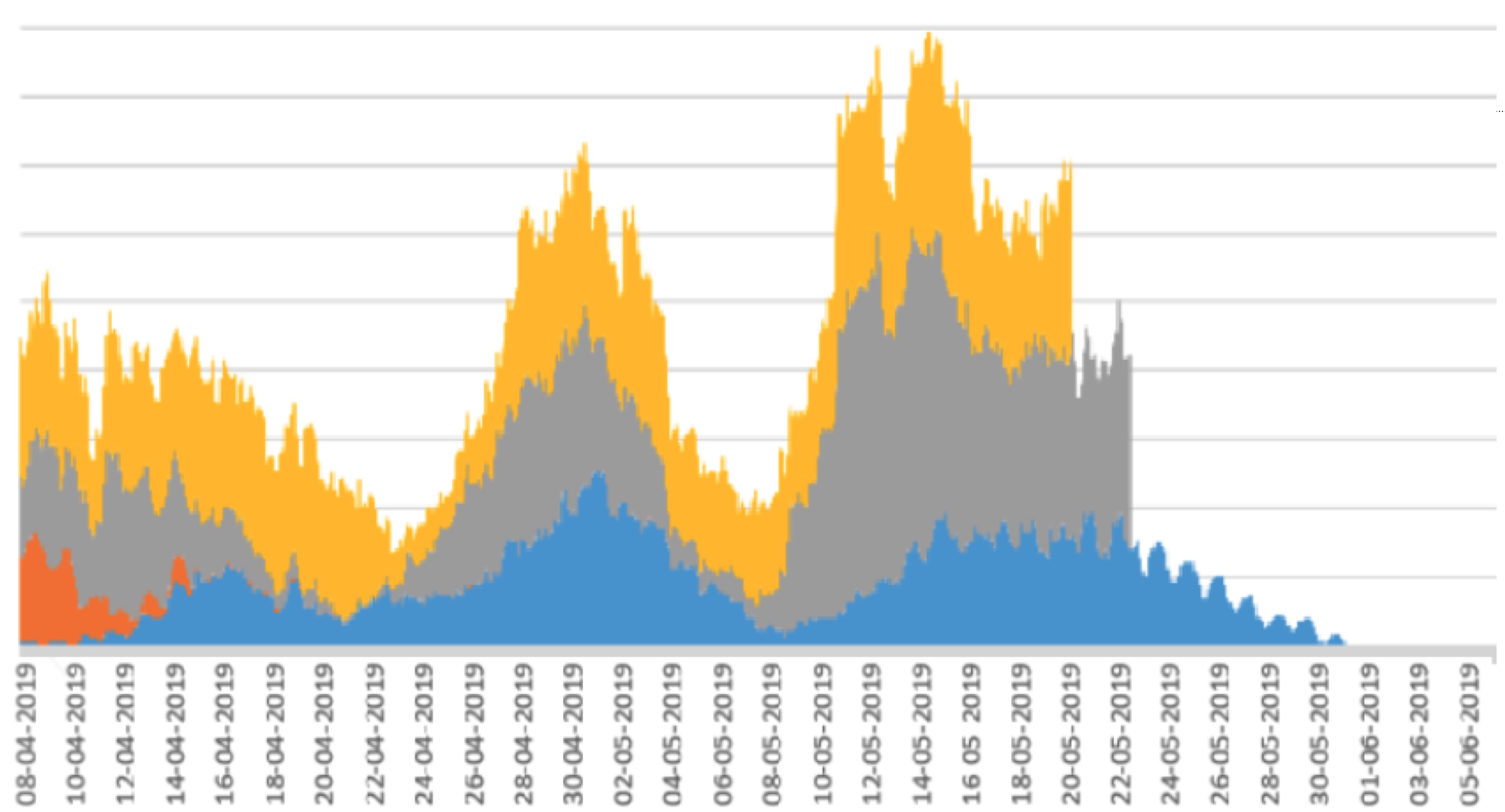}
\caption{Stacked line graph for aligned periodicity intensities for calves 16--19. Each colour represents a different~calf. \label{fig:stack5}}
\end{figure}

Stacked line graphs are useful for visualising the impact {of some event} across a set of variables. In~the case {of the work reported here, the variables are} a set of periodicity intensity values. As~shown in~\cite{4658136}, {stacked line graphs} are known to present an illusion of peaks and troughs which might be caused by {only} a subset of the subjects, calves in our case, rather than across all {of} the subjects. This means that the overall stacked line graph might be artificially boosted or deflated by peaks or troughs from a small number of calves. To~assess this, we divided the main {\deleted{data}} from 19 calves into random sub-{\deleted{samples}groups} and plotted a stacked line graph for each sub-group.  The~rationale for this is that if the stacked line graphs for the sub-groups show the same shape as the {stacked line} graph for the whole herd, then there are no {\deleted{artificical}artificial} boosts or troughs from a sub-group, and the~pattern is more or less applicable to all calves in the herd.  The~stacked line graphs for the sub-groups are shown in Figures~\ref{fig:stack1}--\ref{fig:stack5}.

{The graphs in} Figures~\ref{fig:stack1}--\ref{fig:stack5} show a remarkable consistency in their overall shape, thus illustrating that the majority of  calves show {synchronous} upwards and downwards trends in their periodicity intensities{\deleted{at the same time}}. 
These illustrate that the peaks and troughs  are not artificially boosted by the peaks and troughs of a sub-group of calves. With~the sub-groups and the whole herd, the~same pattern of three peaks is seen as in Figure~\ref{fig:stack} with the middle peak of the periodicity intensity as the highest  also observed in each sub-group. 
From this, we infer that there were likely to be  external events that affected the herd  as a whole. {\deleted{In this case,}While} we were unable to determine the cause of this impact, {\deleted{however}} it does suggest that  periodicity intensity measurements are useful as indicators of the management of environmental factors that may impact on the welfare of calves.
In the case of calf 21427 shown earlier in Figure~\ref{fig:calf21427}, this calf's periodicity intensity values do not precisely match the values of the overall herd, although the first and the third peak are present and~distinguishable.

\unskip

\unskip  

\unskip  

\section{Conclusions}

In this paper, we {\deleted{processed}analysed} three-axis accelerometer movement data sampled at \mbox{12.5 Hz}  from 19 calves over a period of approximately 8 weeks from birth.  The~magnitude of calf movement in 1 min epochs in any direction was calculated and used to compute and visualise the intensity of the 24 h periodicity at {15 min} intervals as the calves became older.

We observe that the intensity of the 24 h periodicities for individual calves increase and decrease synchronously in a herd of 19 calves. Such patterns are not visible from the raw movement data and only by processing the data in this way are these insights revealed.  Our  confidence in this form of analysis is based on generating the visualisation from randomly selected sub-groups of calves as well as from all calves in the herd for which movement data for the full logging period  is~available. 

The results demonstrate{\deleted{s}} that there are factors which affect the welfare of the herd as a whole as calf welfare, for~which periodicity intensity is an indicator, which rises and falls {\deleted{in tandem}synchronously}.  Possible uses of this form of {\deleted{data}} analysis would be the processing of movement or any other form of activity data in~pseudo-real time in order to detect {early stage}  changes in {individual} calves {\deleted{and in} compared to changes in the rest of the herd. 
{Thus, as the herd reacts to external stimuli and influences, both positive influences such as regular feeding or traumatic influences such as disbudding, an~individual calf’s deviations from those synchronised behaviour changes by the herd can be detected.}
\deleted{Given that} Our \deleted{form of}} analysis appears to be more sensitive to {overall herd and individual calf welfare} than currently used forms of analysis {which are based on movement data alone and are used to classify animal  activity~\cite{s21206816} including feeding and drinking~\cite{arcidiacono2017development}, and~overall welfare~\cite{jukan2017smart}}.  

For future work, we recommend that a larger herd size be used for data gathering with a more detailed recording of the internal and external factors which might cause changes in calf or herd welfare such as weaning, turning out to pasture, disbudding, or~even changes in~weather.

\vspace{6pt} 


\authorcontributions{
%
%
Conceptualisation, C.M., V.R. and A.F.S.; 
methodology, C.M., V.R. and A.F.S.; 
software, M.M. and M.S.; 
formal analysis, C.M. and A.F.S.; 
investigation, C.M., V.R. and A.F.S.; 
data curation, V.R.; 
writing---original draft preparation, M.M. and M.S.; 
writing---review and editing, C.M., V.R. and A.F.S.; 
visualisation, A.F.S.; 
funding acquisition, C.M. and A.F.S. 
All authors have read and agreed to the published version of the manuscript.
}

\funding{This research was supported by a UCD Wellcome Institutional Strategic Support Fund, which was jointly financed by University College Dublin and the SFI-HRB-Wellcome Biomedical Research Partnership (ref 204844/Z/16/Z).
It was also  partly supported by Science Foundation Ireland (SFI) under Grant Number SFI/12/RC/2289\_P2 (Insight SFI Research Centre for Data Analytics), co-funded by the European Regional Development Fund.}

\institutionalreview{These data were collected under an ethical exemption from UCD Animal Research Ethics Committee, approval number~AREC-E-19-46-McAloon.
}

\dataavailability{The data presented in this study are openly available in Figshare at {\url{https://doi.org/10.6084/m9.figshare.20039486}  (accessed on 9 June 2022)}.  
} 


\conflictsofinterest{The authors declare no conflict of~interest.} 



\abbreviations{Abbreviations}{
The following abbreviations are used in this manuscript:\\

\noindent 
\begin{tabular}{@{}ll}
PSD & Power Spectral Density \\
SVM & Signal Vector Magnitude \\
\end{tabular}
}




\begin{adjustwidth}{-\extralength}{0cm}

\reftitle{References}


\bibliography{bibliography.bib}

\end{adjustwidth}
\end{document}